\documentclass[aps,prl,twocolumn,showpacs,amssymb,groupedaddress]{revtex4}
\usepackage{graphicx}

\begin{document}


\title{Swarming and swirling in self-propelled polar granular rods}
\author{Arshad Kudrolli$^{\dag}$, Geoffroy Lumay$^{\dag,\ddag}$, Dmitri Volfson$^{*}$,  Lev S. Tsimring$^*$}

\affiliation{\dag Department of Physics, Clark University, Worcester, MA
01610\\
\ddag GRASP, Physics Department, University of Li\`ege, B-4000 Li\`ege,
Belgium\\
* Institute for Nonlinear Science, University of California, San Diego, La Jolla, CA 92093}

\begin{abstract}
Using experiments with anisotropic vibrated rods and quasi-2D numerical
simulations, we show that shape plays an important role in the
collective dynamics of self-propelled (SP) particles. We demonstrate that
SP rods exhibit local ordering, aggregation at the side
walls, and clustering absent in round SP particles.
Furthermore, we find that at sufficiently strong excitation
SP rods engage in a persistent swirling motion in which the
velocity is strongly correlated with particle orientation.
\end{abstract}




\maketitle


Large-scale structures emerge spontaneously in systems of interacting SP
biological objects such as flocks of birds, schools of fish, amoebae
colonies, as well as in multi-robot swarms~\cite{reynolds87,toner05}.
Chemotaxis and field gradients can lead to non-equilibrium
aggregation~\cite{budrene95}, and hydrodynamic interactions
can cause vortices~\cite{dombro04}. Such observations prompted a discrete-time, discrete-element
model~\cite{vicsek95} where SP point particles (``boids")
align their velocities with the average velocity of other particles
within a certain fixed-size neighborhood. This model predicts a
spontaneous phase transition to a long-range ordered state as the noise
(temperature) of the system is reduced below a critical value, however
the exact nature of the transition is still a matter of debate~\cite{gregoire04}. 
Continuum hydrodynamic-type field models for a population of SP particles have been derived 
either general symmetry arguments~\cite{toner95} or directly from microscopic 
interaction rules~\cite{bertin06}. These models allowed for detailed predictions of the correlation
properties within the ordered state. However, both these models did not explicitly
take into account the finite size and shape of interacting particles.
On the other hand, there have been rapid advances in the theory of
``active nematics", or populations of inelastically interacting rods,
both polar~\cite{liverpool03,aranson05} and apolar~\cite{ramaswamy03,narayan06}.  
These models predict onset of a nematic order when the coupling strength of 
particle density becomes sufficiently high, furthermore, clustering of apolar 
rods can lead to giant density fluctuations. Clustering  of polar rods was 
recently found in numerical simulations~\cite{peruani06}.
On the experimental side, there has also been growing
interest in the nonequilibrium dynamics of driven granular rods.
Symmetric rods in a vibrated container have been shown to form nematic
or tetratic order and under certain conditions exhibit persistent
swirling~\cite{narayan06,galanis06,aranson07}, and giant number 
fluctuations~\cite{narayan07}. At higher density, rods begin to bounce 
on one end and travel in the direction of their tilt due to friction at
the contact between the rod and the substrate~\cite{volfson04}.
Collectively, these rods spontaneously form large scale
vortices~\cite{blair03,aranson03}.

Here, we study experimentally and numerically the collective dynamics of
vibrated {\em polar} granular rods which have a non-symmetrical mass
distribution but retain their shape symmetry.  Such a rod on a vibrated 
surface moves towards the lighter end. When many such rods are placed inside a
vibrated container, for weak excitation they aggregate over time at the
boundaries. When the magnitude of excitation is increased,
aggregation at the boundaries is reduced, and coherent structures are
found in the bulk of the container. In particular, swirls can be
identified in time averaged velocity fields, the flow and the rods are
aligned, and signatures of incipient clustering can be observed. To
augment these results and extend them toward larger system sizes, we
perform numerical simulations using a discrete-element molecular dynamics
algorithm. In particular, we show the importance of particle aspect
ratio and driving fluctuations on the observed pattern formation.

\begin{figure}
\includegraphics[width=0.8
\linewidth]{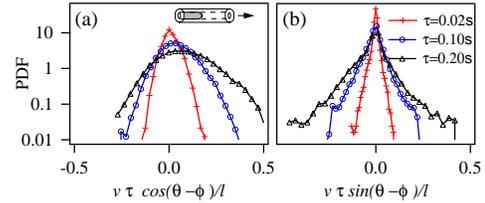}
\caption{The distributions of displacements in a time interval $\tau$
(a) parallel, and (b) perpendicular to axis of the polar rod ($\Gamma
=2$). Inset: Schematic of the polar rod. Arrow shows the direction of net motion. 
} \label{single} \end{figure}

About 10$^3$ polar rods were built using white hollow nylon cylinders of length $l
= 9.5$mm and diameter $d = 4.76$ mm, so the aspect ratio of the rods $A_r$ 
was fixed at 2 (see Fig.~\ref{single}(a), Inset). Solid steel cylinders of length 4.75\,mm and diameter 2.5 mm\,were placed snugly in one end of the nylon tube, which resulted in the center of mass being displaced by 0.1$l$ from the geometrical center of the rod. The total mass of the assembly was $2.20 \times 10^{-4}$~kg. The steel inserts also made the corresponding ends to appear somewhat darker and were used to identify the ``polarity" of the rods. 
The particles were placed on a flat anodized aluminum container of radius $R = 30d$. The container was vertically vibrated using an electromagnetic shaker with sinusoidal waveform at frequency 75 Hz and varied driving acceleration $\Gamma$ (scaled by the gravity acceleration) from 0 to $5$. A digital camera with the spatial resolution of 1000 $\times$ 1000 pixels was used to image the motion of rods inside the container.

First, we studied the motion of a single rod bouncing on the
vibrated plate away from side walls. For $\Gamma  > 1.5$, the rod shows a robust net motion in the
direction of the lighter end of the rod while taking some apparently
random steps in the other directions as well. A movie of the typical
motion is contained in the Supplementary material. By cross-correlating
the intensity distribution of the image of the rod with a mask,
we automated finding the position and the orientation (measured by the angle $\phi$ to a fixed  axis) of the rod in
each frame. By measuring the change in position over time interval $\tau$,
the magnitude of the rod velocity $v$, and its direction $\theta$ with 
respect to a fixed reference were obtained. The probability distribution 
functions (PDF) for the displacement parallel to the rod 
$v \tau \cos(\theta-\phi)$ and perpendicular to the rod $v
\tau \sin(\theta-\phi)$ are plotted in Fig.~\ref{single}(a,b) with $3 \times 10^5$ sets of measurements. While the PDF in the perpendicular direction are centered at zero, the broader PDF in the parallel direction are clearly shifted from
zero, and this shift grows as $\tau$ is increased. The mean and the rms velocity 
increase with $\Gamma$ in our system (see Supplementary materials.) By imaging from the side, we find that rods undergo short collisions with the bottom of the container once every few cycles at random
phases of the cycle (see Supplementary materials.) 

\begin{figure}
\includegraphics[width=0.55\linewidth]{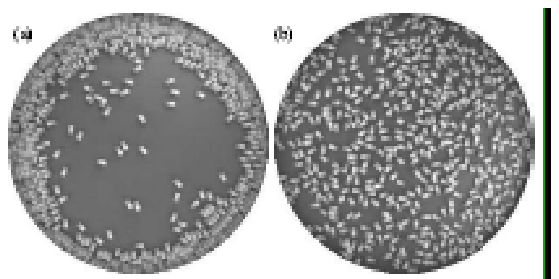}\\
\includegraphics[width=0.7\linewidth]{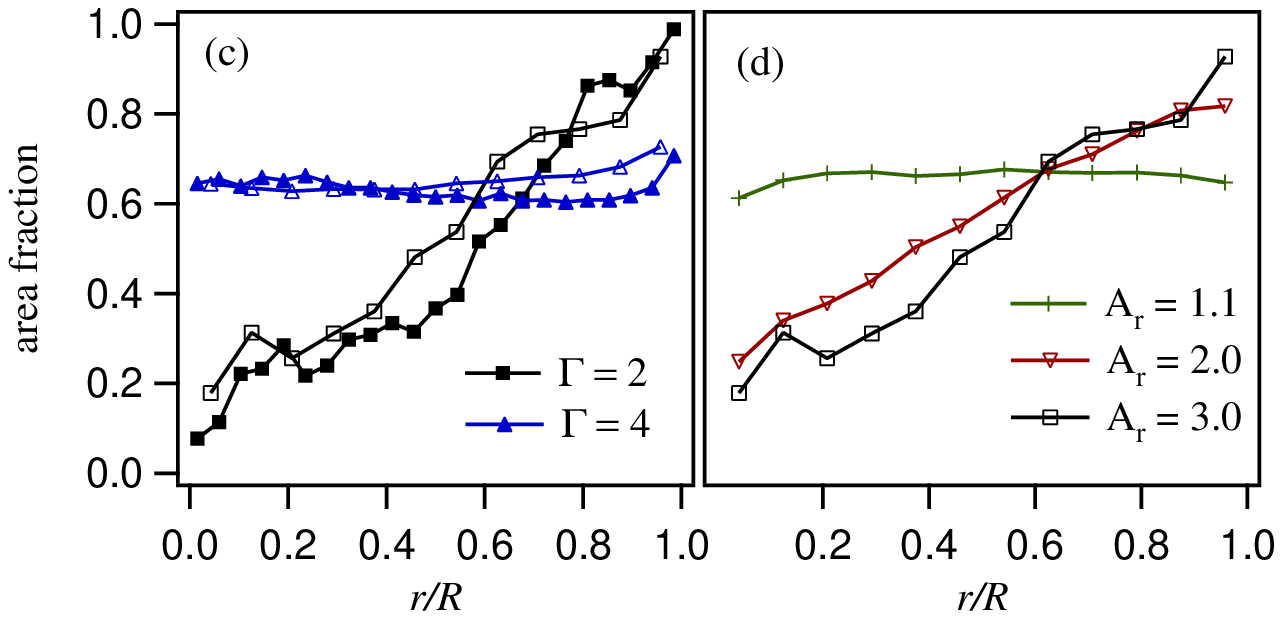}
\caption{(a) Rods migrate and aggregate at the boundaries of a container for modest 
excitations ($N = 500, \Gamma= 2$). (b) Aggregation reduces and a homogeneous distribution 
is observed as excitation is increased. ($N =
500, \Gamma=4$). (c) Area fraction $\rho(r/R)$ as a function of distance $r$
from the center of the container with radius $R$ for $N = 900$ averaged over 100 frames at 10 frames per second, open symbols: the results of numerical simulations
for the same system parameters; (d) Simulations show decrease of clustering as $A_r$ is reduced ($\Gamma=2$).} 
\label{patterns} \end{figure}

The physical mechanism for the observed directed motion in our polar rods can be understood by extending the arguments developed for symmetric rods and dimers~\cite{volfson04,dorbolo05}. During a typical collision of a particle with a horizontal plate, a large but short impulse of frictional force at the contact point causes horizontal particle displacement after the collision. When a symmetrical (apolar) 
rod bounces symmetrically on a vibrated plate, the net displacement after many collisions is absent, but 
for an asymmetric mode of vibration (as in Ref.~\cite{dorbolo05}) or for 
an asymmetric particle (as in the present study), there is a non-zero 
net horizontal motion. In the case of polar rods, since the center of mass is displaced from the geometrical center, the heavy end collides more often with the plate, and the rod on average travels in the direction of the light end. It can be shown that the average horizontal velocity of the rod translation is proportional to the amplitude of the vertical speed of the container, and indeed we observe that the mean velocity increases with $\Gamma$. 

The collective motion of polar rods was studied by placing the rods randomly initially inside the container and then vibrating with various $\Gamma$. (Movies included in Supplementary materials.) For low $\Gamma \sim 2$, rods
were observed to migrate to the boundary of the container and aggregate
in about 30 seconds.  An example is shown in Fig.~\ref{patterns}(a). Not
all rods aggregate at the boundaries, as some rods gradually rotate and escape
from the dense cluster at the boundary back into the middle of the
container. As $\Gamma$ is increased, so do fluctuations, and the
aggregation at the boundaries becomes less and less pronounced. Although
spatio-temporal density inhomogeneities persist, the time-averaged
number density of the polar rods appear more or less uniform across the
cavity for $\Gamma > 3$ (see Fig.~\ref{patterns}(b)). 

Next, we performed ``molecular dynamics" simulations of polar
rod motion and interaction. We did not simulate the details of the
vibrational transport of bouncing rods, but instead assume that the
rods were confined to a horizontal plane. A force acts on each
rod along its (horizontal) axis in the direction of the lighter end.
This force was assumed to be random, with a mean $F$ 
and variance $V$. In addition to the
driving force, we assumed that rods experience velocity-dependent
friction with the substrate and inelastic collisions with other rods. $F$
and $V$ were tuned so the displacement distribution for a single rod fits the
experimental data for a given $\Gamma$. In the numerics, the rods had
a form of spherocylinders, which helped in modeling contact forces.  
The interaction forces among rods were calculated via the interaction between
viscoelastic virtual spheres of diameter $d$ centered at the closest
points between the axes of the spherocylinders~\cite{volfson04}. Normal 
forces were computed using Hertzian spring-dashpot
model, and dynamic Coulomb friction was assumed for tangential forces.
We did not add random forcing in the direction
perpendicular to the axis of the rod,
so the rods could only change their direction
by colliding with the walls or other rods. 

We first performed simulations which matched the experiment both in
terms of number of rods and the system geometry. We used aspect ratio
$A_r=2$ and imposed elastic
boundary conditions on a circle of radius $R_S = 34.2d$~\cite{note_rad}.
For $F =0.25, V=0.16$ which correspond to $\Gamma=2$, we find
that rods tend to aggregate at the boundary in agreement with
experiment. As $F$ and $V$ are increased, the aggregation at the
boundaries diminishes, also in accord with the experimental
observations (see numerical movies in the Supplementary material). 
To illustrate and compare the aggregation of rods in the
experiments and simulations, we plot the  projected rod area fraction
$\rho$ as a function of distance from the center $r$ in
Fig.~\ref{patterns}(c). 

Clustering at the walls is not simply the consequence of inelastic collisions.
Indeed, aggregation doesn't occur at the boundary for small $A_r=1.1$ (Fig.~\ref{patterns}(d)), 
which indicates that particle shape affects
aggregation.  When fluctuations are small (at small $\Gamma$), rods have a much lower
probability of turning around and leaving the wall than spherical
particles, and so they are trapped near the wall for a long time.

\begin{figure}
\includegraphics[width=0.5\linewidth]{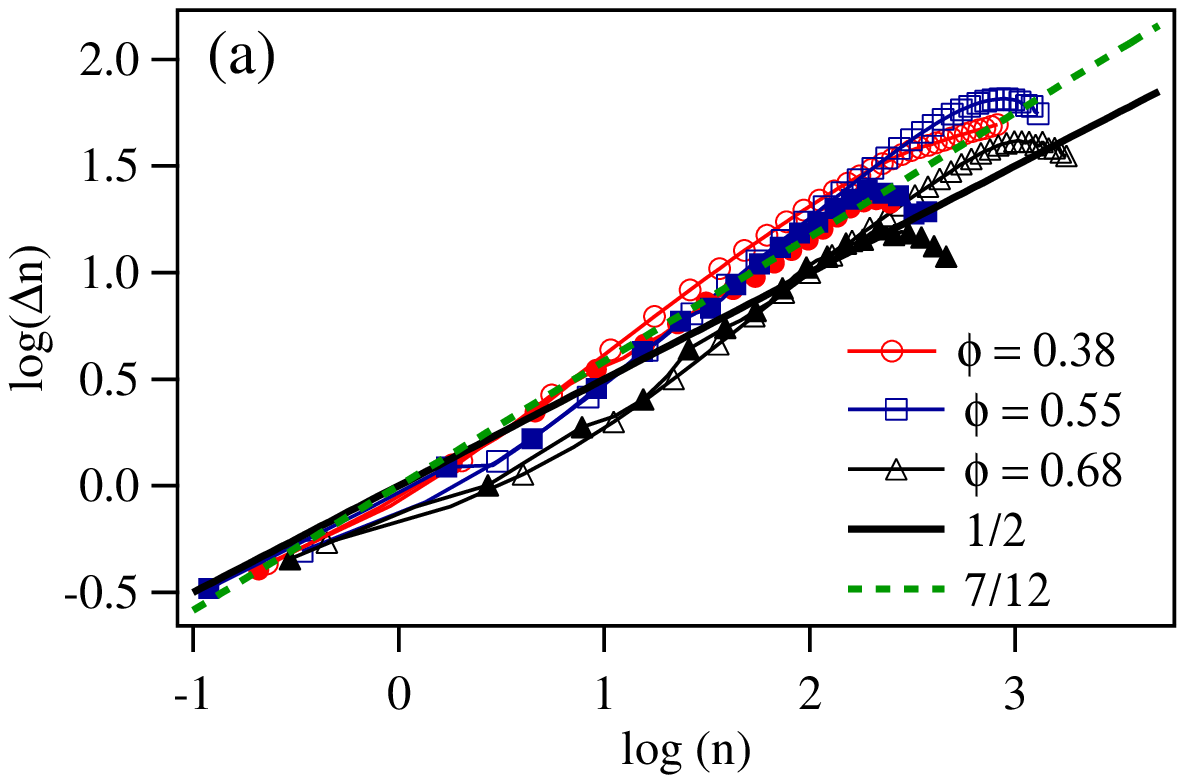}
\includegraphics[width=0.6\linewidth]{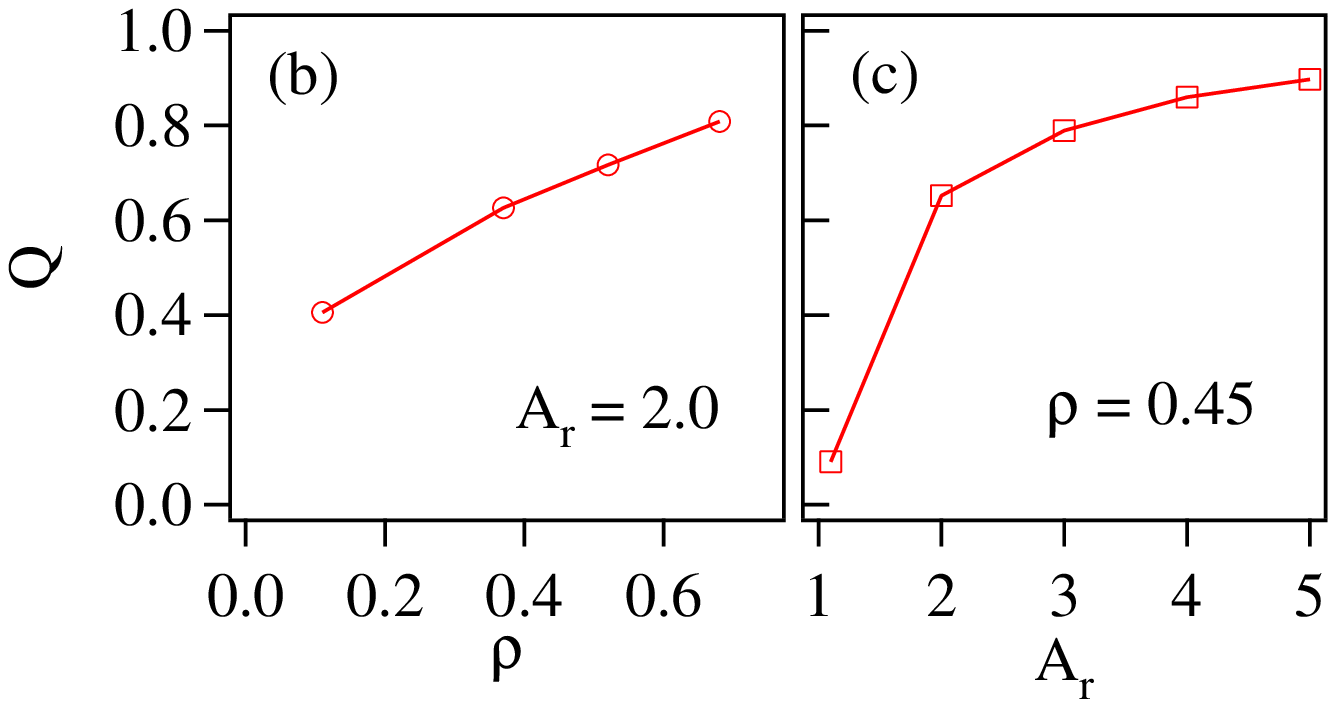}
\caption{(a) The standard deviation of the number of rods $\Delta n$ versus
mean number of rods $n$ inside a circular area at the center of the
container ($\Gamma = 3$). Open symbols correspond to the simulations
for a larger system size $R_L = 2.5 R_S$ but the same density.  We
consistently observe that $\Delta n$  grows {\em faster} than
$\sqrt{n}$, which is a signature of a clustering regime. 
Local orientational order $Q$ for nearest neighbors from simulation data as a 
function of (b) density, and (c) aspect ratio. 
} \label{densfluc} \end{figure}

In order to characterize the density fluctuations inside the container,
we obtain the standard deviation $\Delta n$ and the mean 
$n$ of the number of rods in areas of different sizes by averaging over
many realizations (see Fig.~\ref{densfluc}(a)). The distributions were
obtained by averaging over 1500 frames after the
system reached a statistically stationary regime, and we restricted the
area of measurements to $r/R <0.7$ to minimize boundary effects. 
The data is systematically higher than $\sqrt{n}$. In fact, they are
better described by the slope 7/12 which is predicted by the dynamic XY
model~\cite{toner95} in the nematic state. At very high values of $n$
the  standard deviation drops down, as should be expected since the
number of rods becomes comparable with the total number of
rods in the container. In the numerical simulations which are also
plotted in Fig.~\ref{densfluc}(a), we examine a larger system with $R_L =
2.5R_s$ and almost an order of magnitude greater number of rods.
The deviations from $\sqrt{n}$ are also clearly present in this larger
system. It is interesting to contrast these results with
``giant" ($\Delta n\sim n$) fluctuations reported for apolar rods~\cite{ramaswamy03,narayan07}. 
Although rods in our system have apolar shape, they have mass
anisotropy which renders them polar and self-propelled under external
vibration. This polarity appears to destroy the emergence of giant
density fluctuations in agreement with earlier theoretical work~\cite{toner95}.

Although global orientational order is clearly absent in our system, there is a strong evidence of the local 
orientational order at sufficiently high density of rods. We can characterize this ordering by computing a local 
orientational order parameter $Q$ which we define as 
$Q=\langle\cos2\Theta\rangle$ where $\Theta$ is the angle between directors 
of a rod and its nearest neighbor and brackets indicate averaging over all 
the rods in the container and time (see Fig.~\ref{densfluc}bc). Parameter $Q$ is similar to the local 
orientational order parameter $S$ introduced in~\cite{gregoire04} and shows significant 
local orientational order present in our system at high enough $\rho$ and $A_r$. 

\begin{figure}
\includegraphics[width=0.5\linewidth]{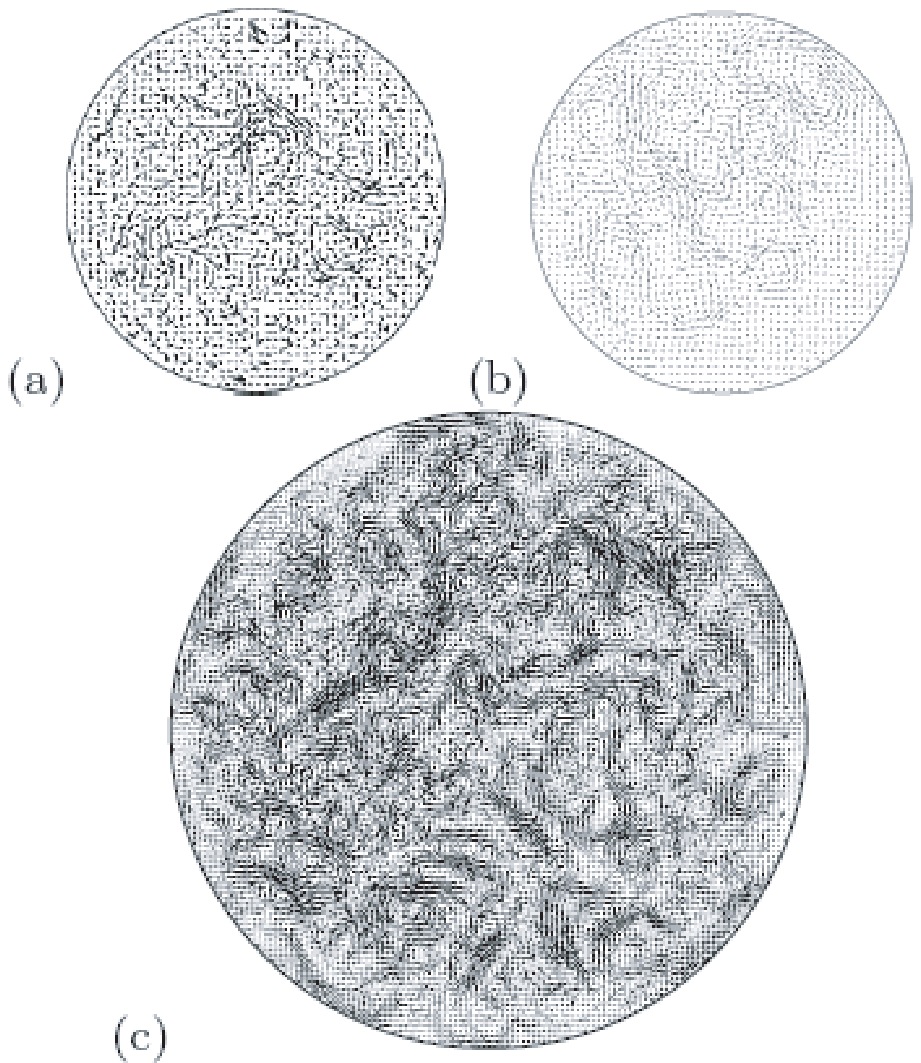}\\
\includegraphics[width=0.4\linewidth]{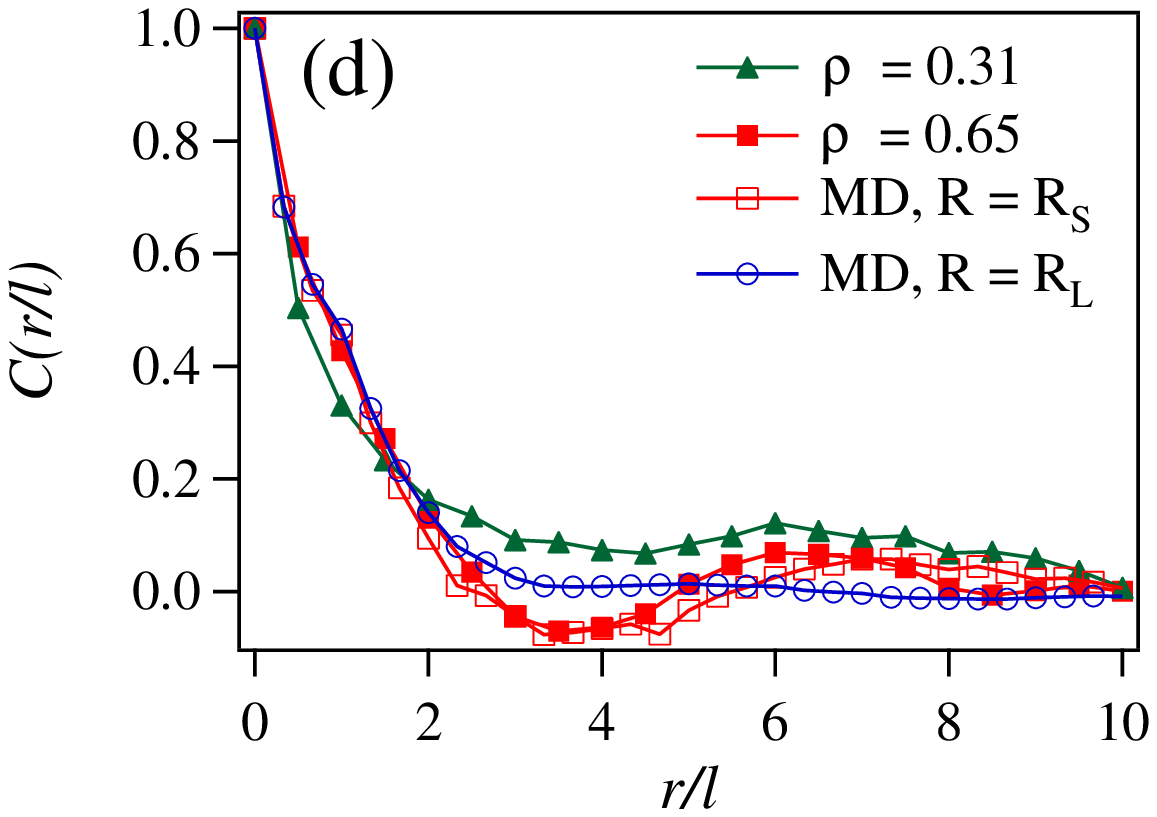}
\includegraphics[width=0.4\linewidth]{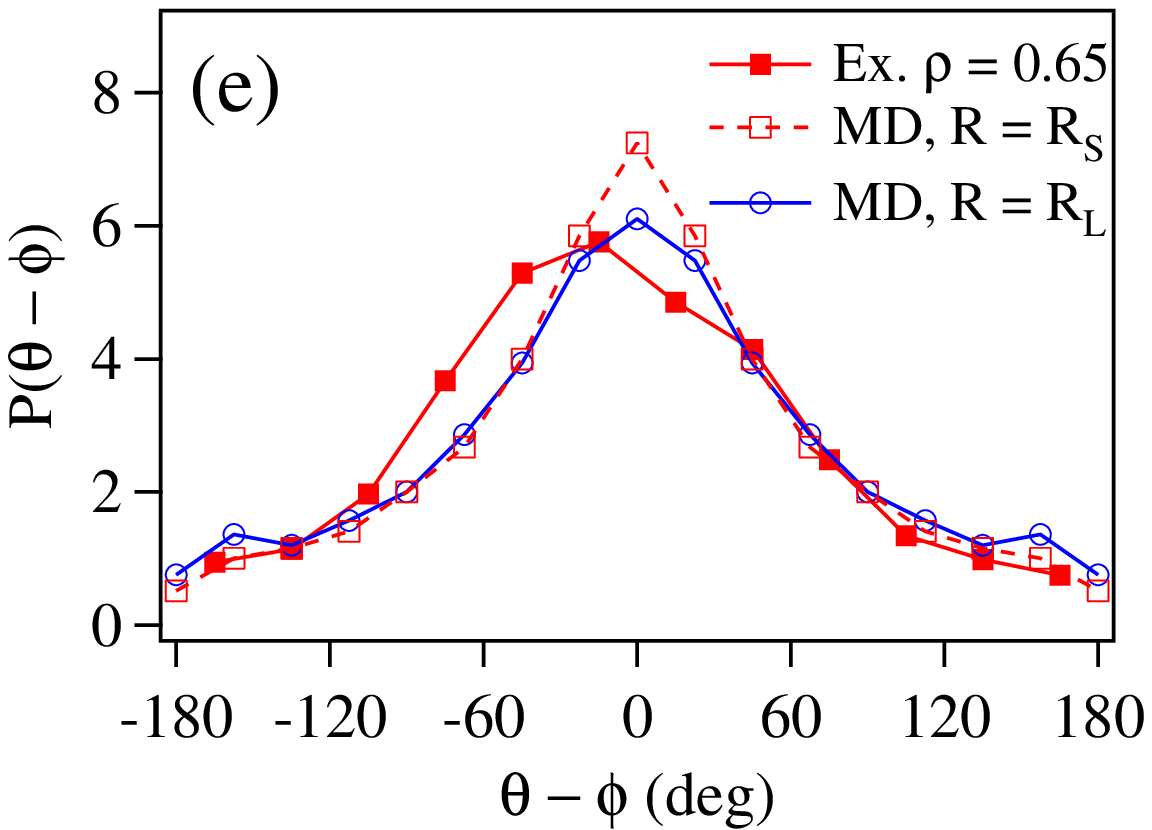}
\caption{Swirling in time averaged velocity obtained by computing particle displacements after $\tau =5$s: (a) experiment ($\Gamma =3 N = 900$), (b) numerics ($F_0 = 1.0, \rho = 0.68$), (c) Example of swirls observed in a larger numerical system ($N=5500$, $R_L=2.5\,R_S$). 
(d) Spatial velocity correlation function $C(r)$ as a function of 
distance between two rods ($\Gamma = 3, \rho =0.31, \rho = 0.65$); results of simulations are shown for small ($R_S=34.2d$) 
and large $R_L=2.5R_S$ system sizes and for the same density $\rho=0.68$.
(e) The distribution of the angle between rod orientation and its velocity.} 
\label{vfield} \end{figure}

Collective motion of rods in the container is masked by the strong
random fluctuations, especially at high $\Gamma$. To reduce these
fluctuations, we divided the field of view into $2d \times 2d$ boxes and
averaged the velocity field over the box area and over a $\tau = 5$
second time interval. An example of the obtained velocity field is shown
in Fig.~\ref{vfield}(a). This procedure reveals a number of streams and
swirls. Numerical simulations for similar parameters also show
swirl-like structures [see Fig.~\ref{vfield}(b)]. 
The coherent structures become more pronounced when the system size
is increased [see Fig.~\ref{vfield}(c)].  These structures are reminiscent of
swirls obtained with apolar particles driven by the 
substrate~\cite{narayan06,aranson07}. However this is not
entirely unexpected since the tensor order parameter for apolar
particles in 2D has only two independent components and the
corresponding order parameter equation can be reduced to a pseudo-vector
form~\cite{aranson07} which is similar to a vector order parameter
equation for polar systems~\cite{aranson03,aranson05,aranson07a}.

To quantify the structure of swirls, we plot in Fig.~\ref{vfield}(d) the
spatial velocity correlation function $C(r) = \langle {\bf v}_1 \cdot {\bf
v_2}\rangle /\langle |{\bf v}_1| |{\bf v}_2|\rangle $ for a rod with velocity ${\bf v}_1$ and
a rod with velocity ${\bf v}_2$ separated by distance $r$.
The correlations decay over a distance of a few rod lengths which
confirms the lack of the long-range order in the system.
However, the structure of the velocity field is strongly correlated with the
orientation of the rods. We computed the distribution of the angle
between the direction of the velocity field $\theta$ in and the mean
orientation within a ($2d\times 2d$) box both in experiment
and numerical simulations, see Fig.~\ref{vfield}(e). As seen in 
Fig.~\ref{vfield}(e), there is a significant maximum of this distribution at angle 0,
which indicates that rods predominantly move along their axes. 

In summary, we have studied the collective dynamics of ``self-propelled" polar rods with experiments and numerical simulations. The phenomenology differs {\em qualitatively} from that of collective motion  of both point-like self-propelled particles~\cite{gregoire04} (which show no tendency to aggregate near the walls and get involved in system-size collective motion) and apolar rods~\cite{ramaswamy03,narayan06,narayan07} (which exhibit giant density fluctuations). We observe aggregation of rods at the boundaries because of the inability of rods to turn around and escape for high enough density under low noise conditions. As vibration strength and thus noise is increased, the aggregation reduces and a uniformly distributed state displaying local orientational order and swirls are observed. 
We observe greater than $\sqrt{n}$ density fluctuations which are in a qualitative agreement with model~\cite{toner95}, but this agreement should not be over-emphasized since the model is directly applicable to a nematic regime. In our opinion, the observed deviation from the $\sqrt{n}$ regime for interacting polar rods is not accounted for by existing models and deserves further study.  In conclusion, our findings elucidate 
an important and interesting interplay between the shape and the directed motion in {\em realistic} self-propelled rods which affects the phenomenology of their collective dynamics. 


\begin{acknowledgments}
The work was supported by the National Science Foundation under Grant
No. DMR-0605664 and the U.S. Department of Energy under Grant No.
DE-FG02-04ER46135.
\end{acknowledgments}



\begin{thebibliography}{0}
\bibitem{note_rad} $R$ in the simulations was chosen
slightly bigger than in experiments to account for the different shapes
of particles in order to match the areal density of rods.
\end{thebibliography}

{\bf Supplementary materials}:\\ http://physics.clarku.edu/$\sim$akudrolli/polarods
\end{document}